\documentclass[
prc,%
10pt,%
final,%
notitlepage,%
oneside,%
twocolumn,%
nobibnotes,%
nofootinbib,
superscriptaddress,%
floatfix,%
floatfix,%
showkeys,%
showpacs]%
{revtex4}
\usepackage{color}
\usepackage{amsfonts}
\usepackage{amsbsy}
\usepackage{mathrsfs}
\usepackage{graphicx}
\def\lsim{\mathrel{\rlap{
\lower4pt\hbox{\hskip-3pt$\sim$}}
    \raise1pt\hbox{$<$}}}     
\def\gsim{\mathrel{\rlap{
\lower4pt\hbox{\hskip-3pt$\sim$}}
    \raise1pt\hbox{$>$}}}     
\def\scr#1{\mbox{\scriptsize #1}}
\begin{document}
\title{
	
Estimates of the baryon densities attainable in heavy-ion collisions 
from the beam energy scan program} 
\author{Yu. B. Ivanov}\thanks{e-mail: Y.Ivanov@gsi.de}
\affiliation{National Research Centre "Kurchatov Institute", 123182 Moscow, Russia} 
\affiliation{National Research Nuclear University "MEPhI",   
Moscow 115409, Russia}
\affiliation{Bogoliubov Laboratory of Theoretical Physics, JINR, Dubna 141980, Russia}
\author{A. A. Soldatov}
\affiliation{National Research Nuclear University "MEPhI",   
Moscow 115409, Russia}
\begin{abstract}
The baryon and energy densities 
attained in fragmentation regions in central Au+Au collisions 
in the energy range of 
the Beam Energy Scan (BES) program at the Relativistic Heavy-Ion Collider (RHIC)
are estimated within the model of the three-fluid dynamics. 
It is shown that 
a considerable part of the baryon charge is stopped in the central fireball. 
Even at 39 GeV, approximately 70\% of the total baryon charge turns out to be 
stopped. 
The fraction of this stopped baryon charge decreases  with 
collision energy rise, from 100\% at  7.7 GeV to $\sim$40\% at 62 GeV. 
The highest initial baryon densities of the equilibrated matter, $n_B/n_0 \approx$ 10, 
are reached in the central region of colliding nuclei at $\sqrt{s_{NN}}=$ 20--40 GeV. 
These highest densities develop up to quite moderate freeze-out baryon densities 
at the midrapidity because the matter of the central fireball  
is pushed out to fragmentation regions by one-dimensional expansion.
Therefore, consequences of these high initial baryon densities 
can be observed only in the fragmentation regions of colliding nuclei in 
AFTER@LHC experiments in the fixed-target mode. 
\end{abstract}
\maketitle

\section{Introduction}

At ultra-relativistic  energies the colliding nuclei pass through
each other, compressing and depositing energy in each other,  
rather than mutually stopping as at lower energies. 
The net-baryon charge remains concentrated in the fragmentation regions, 
which are well separated in the configuration and momentum space from the midrapidity fireball.
Therefore, it is generally accepted that the maximal baryon density is achieved in heavy-ion 
collisions at moderately high energies of the  Nuclotron-based Ion Collider fAcility (NICA) 
in Dubna \cite{Kekelidze:2017ghu} and the Facility for Antiproton and Ion Research (FAIR) 
in Darmstadt \cite{Ablyazimov:2017guv}. 
Analysis of midrapidity hadron yields within the statistical model  \cite{Randrup:2006nr,Randrup:2009ch}
supports this viewpoint. 

However, analysis \cite{3FD-bulk-RHIC}
of bulk observables recently measured by the STAR collaboration \cite{Adamczyk:2017iwn} 
in the BES-RHIC energy range 
indicated a high degree of stopping of the baryon matter 
in the central region of colliding nuclei even at the collision energy of $\sqrt{s_{NN}}=$ 39 GeV
\cite{Ivanov:2017xee}. 
The analysis was performed within model of three-fluid dynamics (3FD) \cite{3FD}.
This finding suggests that transition from complete stopping of the baryon matter 
to the asymptotic transparency at ultra-relativistic  energies is quite graduate. 
The stopped equilibrated baryon matter is formed even at BES-RHIC energies. 
Only its observable consequences are manifested in fragmentation regions of 
colliding nuclei rather than in the midrapidity as at NICA-FAIR energies. 
The stopped baryon matter produced at $\sqrt{s_{NN}}=$ 39 GeV
is pushed out to fragmentation regions because of 
its almost one-dimensional (1D) expansion at later stages of the reaction \cite{Ivanov:2017xee}.

Properties of the baryon-rich fragmentation regions (i.e. the fragmentation fireballs)
produced in central  
heavy-ion collisions were discussed long ago 
\cite{Anishetty:1980zp,Csernai:1984qh,Gyulassy:1986fk,Frankfurt:2002js,Mishustin:2001ib,Bass:2002vm,Mishustin:2006wd,%
MehtarTani:2009dv}.  
Recently the theoretical considerations 
on the internal properties of the fragmentation fireballs
were updated in Ref.
\cite{Li:2016wzh} based on the McLerran-Venugopalan model \cite{McLerran:1993ka}.
%
The  BES-RHIC energies are too low for applicability
of the McLerran-Venugopalan model \cite{Li:2016wzh}. Therefore, phenomenological approaches are required. 
In Ref. \cite{Ivanov:2017xee} the baryon and energy densities 
attained in the fragmentation regions at the collision energy of 39 GeV
were  estimated within the 3FD  model  \cite{3FD}. 

The properties of the fragmentation fireballs in heavy-ion collisions at energies $\sqrt{s_{NN}}<$ 18 GeV 
can be and have already been studied at the Super Proton Synchrotron (SPS) at CERN. 
Recent proposal \cite{Brodsky:2012vg} to perform experiments at the Large Hadron Collider (LHC) at CERN in the
fixed-target mode (AFTER@LHC), if it will be realized, will extend this range to higher collision energies.  
The LHC beam of lead ions
interacting on a fixed target would provide an opportunity
to carry out measurements in the kinematical 
range of the target fragmentation region
at collision energies up to 2.76 GeV per nucleon which is 
equivalent to $\sqrt{s_{NN}}=$ 72 GeV in terms of the center-of-mass 
energy.

In the present paper we estimate the baryon and energy densities 
attained in the central and fragmentation regions in heavy-ion collisions in the BES-RHIC
energy range rather than at the single energy as in Ref. \cite{Ivanov:2017xee}.   
The calculations are done within the 3FD model \cite{3FD,Ivanov:2013wha}  
that is quite successful in reproducing 
the bulk observables \cite{3FD-bulk-RHIC}, 
the elliptic \cite{Ivanov:2014zqa}  
and,  though imperfect, directed flow \cite{Konchakovski:2014gda}
in the midrapidity region at the BES-RHIC energies.

\section{The 3FD model}
\label{Model}

Unlike the conventional hydrodynamics, where local
instantaneous stopping of the projectile and target matter is
assumed, the 3FD description \cite{3FD} takes into account
a finite stopping power resulting in a counterstreaming
regime of leading baryon-rich matter. This generally
nonequilibrium regime of the baryon-rich matter
is modeled by two interpenetrating baryon-rich fluids 
initially associated with constituent nucleons of the projectile
(p) and target (t) nuclei. In addition, newly produced particles
are attributed to a fireball (f) fluid.
Each of these fluids is governed by conventional hydrodynamic equations 
coupled by friction terms in the right-hand sides of the Euler equations. 
These friction terms describe energy--momentum loss of the 
baryon-rich fluids. 
A part of this
loss is transformed into thermal excitation of these fluids, while another part 
gives rise to the particle production into the fireball fluid.
The produced fireball fluid in its turn also interacts with other fluids by means 
of friction forces. Thus, 
the 3FD approximation is a minimal way to simulate the early-stage nonequilibrium at
high collision energies. The 3FD model describes the nuclear collision from the stage of the incident cold nuclei
approaching each other to the final freeze-out stage.

A hybrid model based on similar concepts was recently developed in Ref. \cite{Shen:2017bsr}. 
Unlike the 3FD, the hybrid hydrodynamics \cite{Shen:2017bsr} deals with a single equilibrated fluid that however 
does not involve all the matter of colliding nuclei. Therefore, this hybrid hydrodynamics contains source 
therms describing gain of the equilibrated matter in the course of the collision. This is similar to the 
production of the f-fluid in the 3FD, while the baryon-rich matter in the 3FD is described by two separate 
(p and t) 
fluids which are locally unified (i.e. equilibrated) into a single baryon-rich fluid 
only when they are sufficiently decelerated.

The physical input of the present 3FD calculations is described in
Ref.~\cite{Ivanov:2013wha}. 
The simulations in 
\cite{Ivanov:2013wha,3FD-bulk-RHIC,Ivanov:2014zqa,Konchakovski:2014gda} 
were performed with different 
equations of state (EoS's): a purely hadronic EoS \cite{gasEOS}  
and two versions of the EoS involving the   deconfinement
 transition \cite{Toneev06}, i.e. a first-order phase transition  
and a smooth crossover one. In the present paper we demonstrate results with 
only the first-order-transition and crossover EoS's as the most successful in reproduction of various 
observables at the BES-RHIC energies \cite{3FD-bulk-RHIC,Ivanov:2014zqa,Konchakovski:2014gda}. 

\begin{figure}[!h]
\includegraphics[width=6.cm]{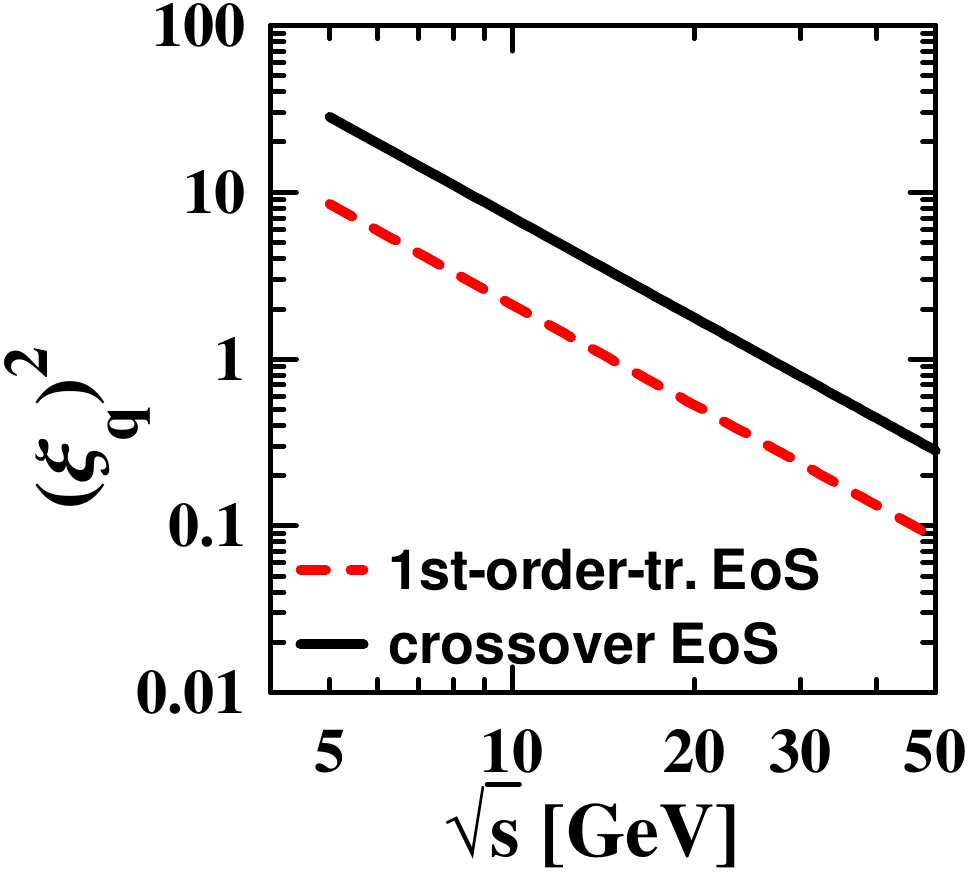}
 \caption{
The ratio of the friction in the QGP to  
that in the hadronic phase [$(\xi_q)^2$ in notation of Refs. \cite{3FD,Ivanov:2013wha}, 
see  Eqs. (14) and (16) in Ref. \cite{Ivanov:2013wha}]  
as a function of $\sqrt{s}$, i.e.
the center-of-mass energy of two 
nucleons belonging to the counterstreaming fluids, that locally characterizes 
the  relative velocity of these counterstreaming fluids. 
Fits for the first-order-transition and crossover EoS's are presented.
}
\label{fig0}
\end{figure}

Friction forces between fluids are key constituents of the model that 
determine dynamics of the nuclear collision. 
In the hadronic phase the friction forces, estimated in Ref. \cite{Sat90},  
are used in simulations. 
There are no theoretical estimates of
the friction in the quark-gluon phase (QGP) so far.
Therefore, the phenomenological friction in the QGP 
was fitted to reproduce the baryon
stopping at high incident energies within the deconfinement
scenarios as it is described in  Ref. \cite{Ivanov:2013wha} in detail.
This fit resulted in the friction in the QGP that strongly differs from 
that in the hadronic phase estimated in Ref. \cite{Sat90}. 
At low relative velocities of the interpenetrating baryon-rich fluids 
($\sqrt{s}<$ 20--30 GeV, depending on the EoS)\footnote{ 
$\sqrt{s}$ is a running variable locally characterizing 
this relative velocity in terms of the center-of-mass energy of two 
nucleons belonging to these  counterstreaming fluids. This variable 
changes in time and space  \cite{3FD,Ivanov:2013wha}.}
the QGP friction considerably exceeds the hadronic one, while at high relative velocities
($\sqrt{s}>$ 20--30 GeV) the QGP friction becomes lower than the hadronic one. 
This is illustrated in Fig. \ref{fig0}. 
The weak friction at $\sqrt{s}>$ 20--30 GeV does not actually mean high transparency 
of the counterstreaming fluids. An efficient stopping of the baryon-rich fluids takes place 
here because of the friction with the f-fluid that is quite dense at these energies. 
Transition from the hadronic to QGP friction is gradual because even the first-order transition 
proceeds through the mixed phase and gradually starts from the central region of the colliding nuclei.

\begin{figure}[htb]
\includegraphics[width=6.cm]{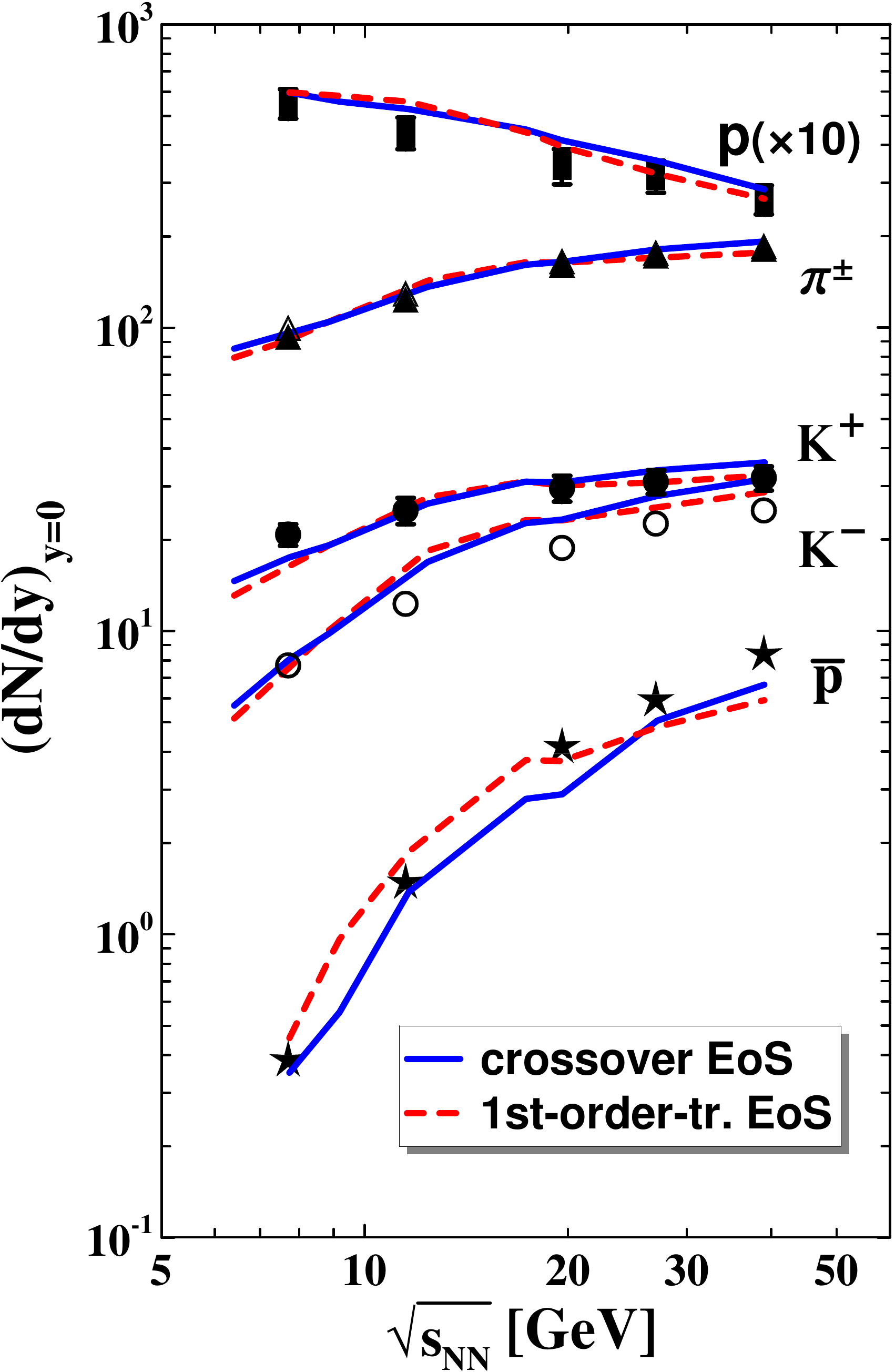}
 \caption{
Rapidity densities $dN/dy$ in the midrapidity of various particles 
produced in central ($b=$ 2 fm) Au+Au collisions 
as functions of the center-of-mass energy
 of colliding nuclei.  3FD calculations are done with  two considered EoS's. 
Experimental data for centrality 0-5\% are from STAR collaboration \cite{Adamczyk:2017iwn}.
}
\label{fig1}
\end{figure}

Figure \ref{fig1} demonstrates the reproduction of
midrapidity densities, $dN/dy$, of various particles 
produced in central (impact parameter $b=$ 2 fm) Au+Au collisions
at the BES-RHIC energies. Experimental data are 
taken from Ref. \cite{Adamczyk:2017iwn}. 
A more detailed comparison with the STAR data on bulk observables \cite{Adamczyk:2017iwn}
can be found in Ref. \cite{3FD-bulk-RHIC}. 
The BES RHIC energy range partially overlaps with that of 
 the SPS, where data are available in 
a wide range of rapidities.  
At SPS energies the 3FD model reproduces data in this wide rapidity range 
\cite{Ivanov:2013wha,Ivanov:2013yqa} rather than only at the midrapidity. 

A numerical "particle-in-cell" scheme is used in the
  simulations, see Ref.~\cite{3FD} and references therein
  for more details. 
  The accuracy requirements result 
in a high computation memory consumption rapidly increasing with the collision
energy, approximately as $\propto s_{NN}$, and a long computation time $\propto (s_{NN})^{3/2}$. 
The reason is the Lorentz contraction of incident
nuclei. 
The grid in the beam, Lorentz-contracted direction ($z$) 
should be fine enough
for a reasonable description of the longitudinal gradients of the matter.
From the practical point of
view, it is desirable to have not less than 40 cells on the
Lorentz-contracted nuclear diameter.
On the other hand, 
to minimize the numerical diffusion in the computational scheme, 
an equal-step grid in all directions ($\Delta x : \Delta y : \Delta z =
1 : 1 : 1$) should be taken, in spite of Lorentz-contraction of the colliding 
nuclei, which is quite strong at high energies. 
This choice makes the scheme isotropic with respect to the numerical
diffusion. However, 
it makes the grid too fine in the transverse directions
and thus results in high memory consumption. 
The need
for the equal-step grid in all directions for relativistic
hydrodynamic computations within the conventional one-fluid
model was pointed out in Ref. \cite{Waldhauser:1992xf}. As it was
demonstrated there, the matter transport becomes even
acausal if this condition is strongly violated. 
Precisely these numerical constraints and available computation 
resources do not allow us to perform simulations for energies 
above 39 GeV. However, in the present paper we occasionally demonstrate 
results at 62 GeV, though they are not quite reliable numerically: 
Only 21 cells on the Lorentz-contracted nuclear diameter were taken.

\section{Evolution of the matter in central region of colliding nuclei}
\label{central region}

\begin{figure}[htb]
\includegraphics[width=8.5cm]{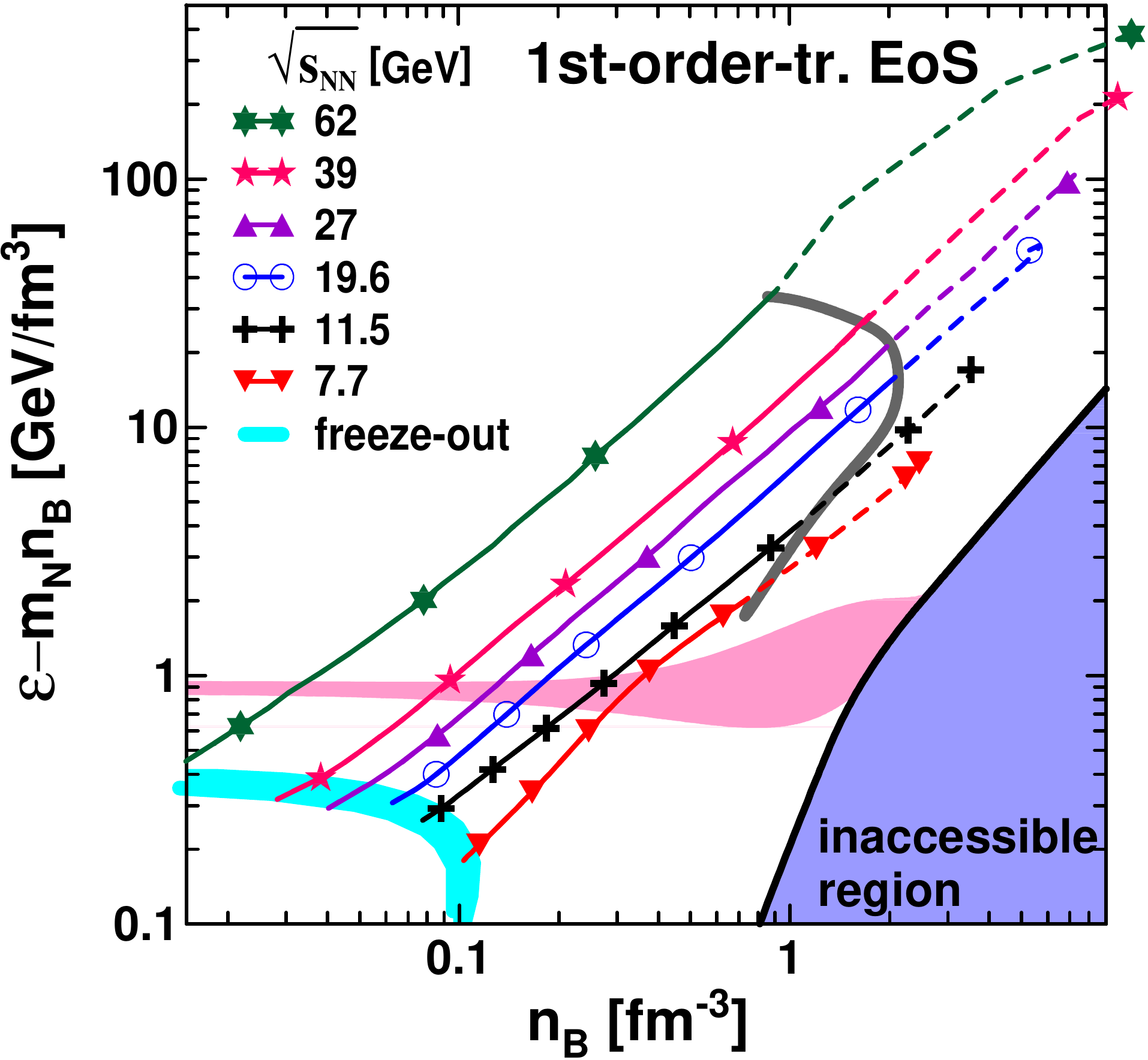}
 \caption{
Dynamical trajectories of the matter in the central region of the
colliding Au+Au nuclei  
($b=$ 2 fm) at BES-RHIC energies. 
The trajectories are plotted in terms
of baryon density ($n_B$) and 
the energy density minus $n_B m_N$, ($\varepsilon - m_N n_B$),
where $m_N$ is the nucleon mass. 
Only expansion stages of the
evolution are displayed.  
Symbols on the trajectories indicate the time rate of the evolution:
time span between marks is 1 fm/c. 
Evolution of the equilibrated matter is displayed by solid lines, while 
the stage before the equilibration - by dashed lines. 
The trajectories are presented 
for the first-order-transition EoS. 
The mixed phase is displayed by 
the shadowed region. 
Inaccessible region is restricted by $\varepsilon(n_B,T=0)-m_N n_B$ from above. 
The bold gray line displays the boundary of initial equilibration.  
The freeze-out aria \cite{Randrup:2006nr,Randrup:2009ch} is displayed by 
the cyan shadowed region. 
}
\label{fig1a}
\end{figure}

Figure \ref{fig1a} presents the dynamics of nuclear collisions at BES-RHIC energies 
in the central region of colliding nuclei. 
Similarly to Ref. \cite{Randrup07}, 
the figure displays dynamical 
trajectories of the matter in the central box placed around the
origin ${\bf r}=(0,0,0)$ in the frame of equal velocities of
colliding nuclei:  $|x|\leq$ 2 fm,  $|y|\leq$ 2 fm and $|z|\leq$
$\gamma_{cm}$ 2 fm, where $z$ is the direction of the beam
and $\gamma_{cm}$ is the Lorentz
factor associated with the initial nuclear motion in the c.m. frame.  
The size of the box was chosen 
to be small enough to consider the matter in it as a homogeneous
medium. 
%
Only expansion stages of the evolution are displayed. 
Evolution proceeds from the top point of the trajectory downwards.

The trajectories are presented in terms of baryon density $n_B$ and the energy density $\varepsilon$. 
These quantities require definitions in view of the three-fluid nature of the 3FD model. 
Within the 3FD model the system is characterized by three hydrodynamical velocities,  
$u_{\alpha}^{\mu}$ with $\alpha=$ p, t and f, attributed to these fluids. 
We define a collective 4-velocity of the baryon-rich matter associating it 
with the total baryon current,  
   \begin{eqnarray}
   \label{bar-u}
   u^{\mu}_B =  J_{B}^{\mu}/|J_{B}|, 
   \end{eqnarray}
where $J_{B}^{\mu} = n_{\scr p}u_{\scr p}^{\mu}+n_{\scr t}u_{\scr t}^{\mu}$ is the baryon
current defined in terms of proper baryon densities $n_{\alpha}$ and
hydrodynamic 4-velocities $u_{\alpha}^{\mu}$, and 
   \begin{eqnarray}
   \label{nb-prop}
   |J_{B}|= \left(J_{B}^{\mu} J_{B\mu}\right)^{1/2}\equiv n_B
   \end{eqnarray}
is the proper (i.e. in the local rest frame) baryon density of the p and t  fluids. 
The total proper energy density of all three fluids in the local rest frame, i.e. 
where the composed matter is at rest, is defined as follows
\begin{eqnarray}
\label{eps_tot}
\varepsilon = u_\mu T^{\mu\nu} u_\nu. 
\end{eqnarray}
It is defined in terms of the total energy--momentum tensor 
\begin{eqnarray}
\label{T_tot}
T^{\mu\nu} \equiv
T^{\mu\nu}_{\scr p} + T^{\mu\nu}_{\scr t} + T^{\mu\nu}_{\scr f}
\end{eqnarray}
being the sum of conventional hydrodynamical energy--momentum tensors of separate fluids
and the total collective 4-velocity of the matter
\begin{eqnarray}
\label{u-tot}
u^\mu = u_\nu T^{\mu\nu}/(u_\lambda T^{\lambda\nu} u_\nu). 
\end{eqnarray}
Note that definition (\ref{u-tot}) is, in fact, an equation
determining $u^\mu$. In general, this $u^\mu$ does not coincide with 
4-velocities of separate fluids. 
This definition of the collective 4-velocity is in the spirit of the
Landau--Lifshitz approach to viscous relativistic hydrodynamics.

At a given density $n_B$, the zero-temperature
compressional energy, $\varepsilon(n_B,T=0)$, presents a lower bound on
the energy density $\varepsilon$, therefore the accessible region is correspondingly
limited. 
The non-equilibrium stage of the expansion is displayed by dashed lines 
in Fig. \ref{fig1}. 
The criterion of the equilibration is equality of longitudinal  
($P_{\rm{long}}=T_{zz}$)
and transverse 
($P_{\rm{tr}}=[T_{xx}+T_{yy}]/2$)
pressures
%
in the box with the accuracy better than 10\%. 
Here $T_{\mu\nu}$ is the energy--momentum tensor of composed matter (\ref{T_tot}). 
The spatial components of the hydrodynamical four-velocity of the composed matter 
in the considered central box are zero due to symmetry reasons. Therefore, the c.m. frame 
of colliding nuclei coincides with the local rest frame of composed matter in the box. 
Note that the equilibration of the medium was not analyzed in the original paper \cite{Randrup07}.

The trajectories for the first-order-transition  and crossover  
EoS's are very similar, as shown in Ref. \cite{3FD-bulk-RHIC}. 
Therefore, here we present only the trajectories for the first-order-transition EoS.  
The above-mentioned similarity is not because of similarity of these two EoS's. It takes place
because the friction forces in the QGP were independently fitted 
for each EoS in order to reproduce observables in the midrapidity region. 
As an estimate for the top LHC energy in the fixed-target mode
the trajectory for energy of 62 GeV is also presented in spite of not quite
reliable numerics.
%

Comparison of 
the 3FD results in the central box with similar results of Ref. \cite{Bravina:2008ra} 
allows us to reveal the effect of the enhanced friction in the QGP. 
Two models were used in Ref. \cite{Bravina:2008ra} 
to study the equilibration in the central box:  
the Quark-Gluon String Model (QGSM) \cite{Amelin:1989vp}
and the model of the Ultrarelativistic Quantum Molecular Dynamics (UrQMD) \cite{Bass:1998ca}. 
The baryon stopping and hence equilibration are treated in hadronic terms within these models.  
At the collision energy of 7.7 GeV (or 30A GeV in the lab. frame)
the equilibration time in a small box (0.5 fm $\times$ 0.5 fm $\times$ 0.5 fm) 
within the QGSM and UrQMD \cite{Bravina:2008ra} is very similar to the 3FD time. 
We do not compare with the results in the large box 
(5 fm $\times$ 5 fm $\times$ 5 fm) \cite{Bravina:2008ra} 
because it is too large to consider the matter in it as homogeneous medium. 
At the top SPS energy the QGSM and UrQMD equilibration times are noticeably longer 
than the 3FD time at the similar energy of 19.6 GeV. 
Respectively, the higher equilibrium densities are reached within the 3FD simulations. 
This is the effect of the stronger QGP friction 
required for reproduction of the SPS data 
\cite{Ivanov:2013wha,Ivanov:2014zqa,Konchakovski:2014gda}.

In contrast to the conventional scenario at top RHIC and LHC energies, 
the equilibration in the central region at BES-RHIC energies, including 62 GeV, 
is achieved at quite high baryon densities. The bold gray line in Fig. \ref{fig1a}
indicates the boundary of the initial equilibration in the central region. 
In a way it is an analog of the hadronic freeze-out line in Ref. \cite{Randrup:2006nr,Randrup:2009ch} 
which is displayed by a cyan shaded area in the lower left corner of Fig. \ref{fig1a}. The borders of this aria 
correspond to the freeze-out in terms of the hadronic gas EoS without an excluded volume 
\cite{Randrup:2006nr} (the upper boundary) and that with the excluded volume, $c=$ 0.3 fm,  
 \cite{Randrup:2009ch} (the lower boundary). 
Similarly to the hadronic freeze-out line 
the equilibration line also manifests a maximum baryon density attained at the equilibration. 
It occurs at $\sqrt{s_{NN}}\approx$ 20--40 GeV in central Au+Au collisions, while 
for the hadronic freeze-out line \cite{Randrup:2006nr,Randrup:2009ch} -- at $\sqrt{s_{NN}}\approx$ 8 GeV. 
As seen from Fig. \ref{fig1a}, the equilibration line is converted into the freeze-out line
along the displayed trajectories. 
In particular, the highest equilibrated baryon densities evolve to quite moderate 
freeze-out baryon densities because the baryon-rich matter is pushed out to peripheral 
regions by almost 1D hydrodynamic expansion discussed below.

The above features of the evolution in the central box are consequences of interplay of the enhanced friction 
in the QGP (see Fig. \ref{fig0}) and the finite thickness of colliding Au nuclei.
In particular, the maximal equilibrated baryon density  would be lower and attained at low collision energies 
in collisions of lighter nuclei. On the other hand, in collisions of infinitely thick slabs the strength of 
friction affects only the equilibration time, while the maximal equilibrated baryon and energy densities 
would be the same as in the shock-wave scenario and monotonically increase with the collision energy rise. 
The same interplay determines the global evolution displayed in Figs. \ref{fig2} and \ref{fig3} below.

\section{Global evolution of the matter}
\label{Global evolution}

\begin{figure*}[!htb]
\includegraphics[width=17.5cm]{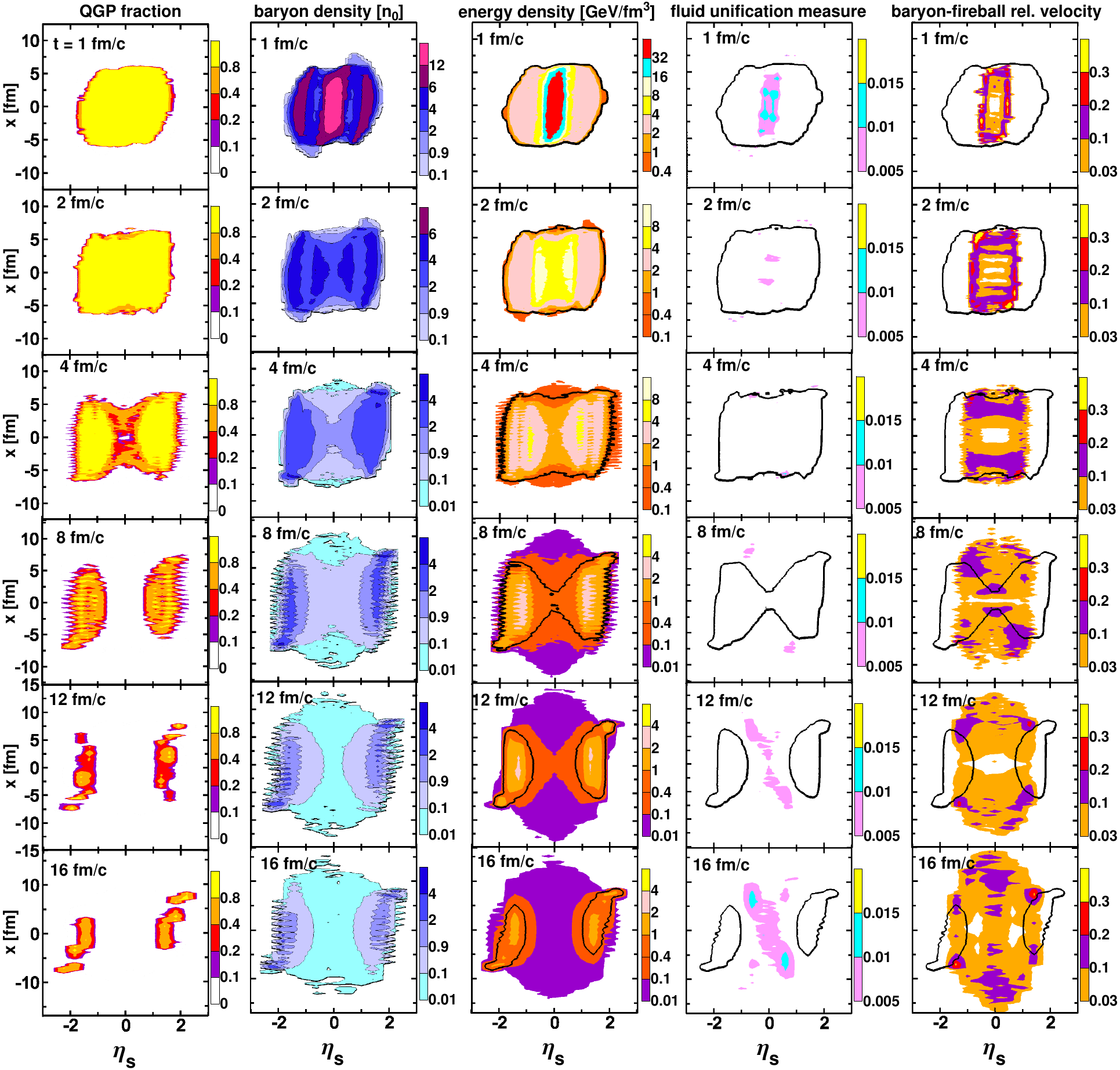}
 \caption{
QGP fraction (first column from the left), the proper baryon density in units of 
the normal nuclear density, $n_0=0.15$ 1/fm$^3$, see Eq. (\ref{nb-prop}) (second column),   
the proper energy density, see Eq. (\ref{eps_tot}) (3d column),  
the baryon-fluid unification measure, see Eq.  (\ref{unification}) (4th column),
the baryon-fireball relative velocity, see Eq.  
(\ref{fb-rel-vel})  (5th column)
in the reaction plain ($x\eta_s$) at various time instants 
in the central ($b=$ 2 fm) Au+Au collision at $\sqrt{s_{NN}}=$ 39 GeV. 
$\eta_s$ is the space-time rapidity along the beam direction, see Eq. (\ref{eta_s}). 
Calculations are done with the first-order-transition EoS.  
The bold contours in the last three columns on the right
display the boundary between the frozen-out matter and still hydrodynamically
evolving matter.  
}
\label{fig2}
\end{figure*}
\begin{figure*}[!tbh]
\includegraphics[width=17.cm]{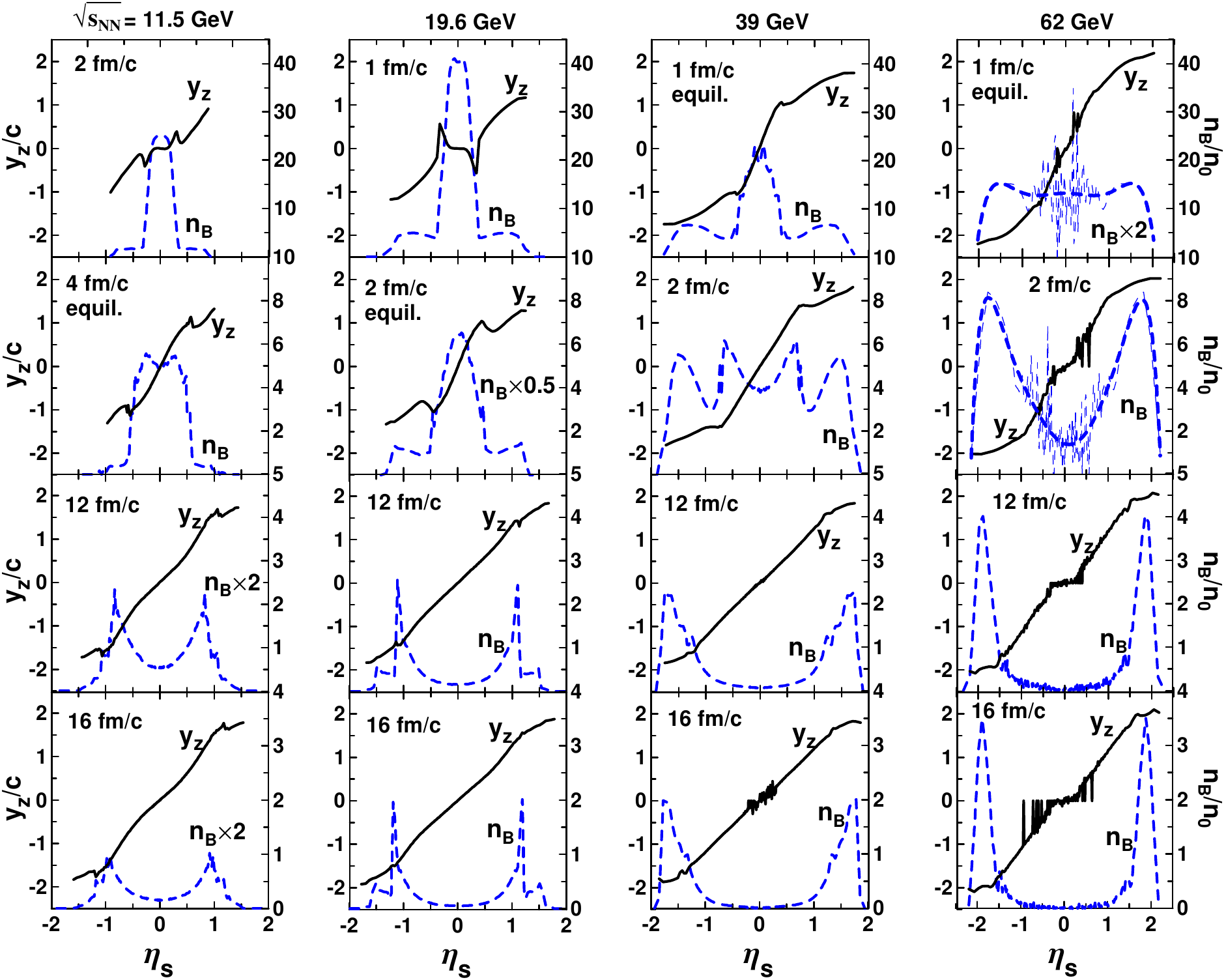}
 \caption{
The proper baryon density ($n_B$) of colliding nuclei in units of 
the normal nuclear density, $n_0=0.15$ 1/fm$^3$ (dashed lines, right scale axis)  
and the longitudinal rapidity ($y_z$) of the matter (solid lines, left scale axis)
 along the beam axis $\eta_s$ ($x=y=0$) 
at various time instants 
in the central ($b=$ 2 fm) Au+Au collision at $\sqrt{s_{NN}}=$ 11.5, 19.6, 39 and 62 GeV.
Calculations are done with the first-order-transition EoS. 
The calculated densities at $\sqrt{s_{NN}}=$ 62 GeV (thin lines) are interpolated by smooth (bold) lines 
when they reveal large numerical fluctuations, see ($t=1$ and 2 fm)-panels. 
The time instants, at which the equilibration is first achieved, 
are marked by the ``equil.'' label. 
}
\label{fig3}
\end{figure*}

Figure \ref{fig2} presents the time evolution of the QGP fraction,    
the proper baryon and energy densities, Eqs. (\ref{nb-prop}) and (\ref{eps_tot}), 
respectively, 
in the reaction plain ($x\eta_s$) of
central ($b=$ 2 fm) Au+Au collision at $\sqrt{s_{NN}}=$ 39 GeV,
where 
   \begin{eqnarray}
   \label{eta_s}
   \eta_s = \frac{1}{2} \ln\left(\frac{t+z}{t-z}\right)
   \end{eqnarray}
is the space-time rapidity and $z$ is the coordinate along the beam direction. 
The advantage of this
longitudinal space-time rapidity is that it is equal to the
kinematic longitudinal rapidity [see Eq. (\ref{y_z})] in the self-similar 1D
expansion of the system.
The figure also displays  
the fluid unification measure  
   \begin{eqnarray}
   \label{unification}
   1-\frac{n_{\scr p}+n_{\scr t}}{n_B}
   \end{eqnarray}
and the baryon-fireball relative velocity  
\begin{eqnarray}
\label{fb-rel-vel}
v_{{\scr f}B} = \sqrt{(u_B\cdot u_{\scr f})^2-1}.  
\end{eqnarray}
The equilibration criterion based on the difference of longitudinal  
and transverse pressures, as it was used in Fig. \ref{fig1a},  
 is not practical outside the central region 
because of nonzero spatial components of the velocity. 
We demonstrate results only for the first-order-transition EoS because 
the crossover scenario gives a very similar picture. 
Moreover, the pattern displayed in Fig. \ref{fig2} for $\sqrt{s_{NN}}=$ 39 GeV
is also representative for $\sqrt{s_{NN}}=$ 19.6 and 27 GeV.

As seen, at $\sqrt{s_{NN}}=$ 39 GeV the baryon-rich fluids are  
mutually stopped and unified already at $t\gsim 1$ fm/c because  
the fluid unification measure  (\ref{unification}) is small: 
it is  less than 0.015 at $t= 1$ fm/c and is practically zero  
inside the freeze-out contour at later instants. This unification measure 
is identically zero, when the p and t  fluids are unified, and  
has a positive value increasing  with the rise of
the relative velocity of the p and t  fluids. 
The baryon-fireball relative velocity (\ref{fb-rel-vel}) is small,  
$v_{{\scr f}B} \lsim 0.2$ at $t \geq$ 1 fm/c. 
This indicates that the system is close 
to the equilibrium. 
We would like to emphasize that this is 
kinetic (i.e. mechanical) rather than chemical equilibrium. For the baryon-rich (p and t) fluids 
the smallness of their local relative velocities is a trigger for their local unification 
into a unified baryon-rich fluid \cite{3FD}, while the f-fluid and the unified baryon-rich fluid 
keep their identity even at small $v_{{\scr f}B}$ and thus do
not provide chemical equilibrium in the composed system.
The f-fluid  is gradually absorbed by the baryon-rich fluid \cite{3FD},  
nevertheless, it survives until the very freeze-out. 
In particular, because of the absence of the chemical equilibrium a unified freeze-out for 
simultaneous description of $p_T$ spectra
and hadron abundances became possible \cite{3FD-bulk-RHIC}.
Below, the term ``equilibration'' is understood precisely in this kinetic sense. 
Note that the above unification/equilibration measures 
are meaningful only within the borders of the freeze-out 
(bold contours in Fig. \ref{fig2}) because the matter 
is frozen out \cite{Russkikh:2006aa} at this boundary and its further evolution has no practical meaning.

As seen from Fig. \ref{fig2}, at $t=$ 1 fm/c
the matter of colliding nuclei has already partially passed though each other
(two narrow bumps of the baryon density near $\eta_s=\pm$ 1 fm)
and partially stopped in the center region (the center bump in $n_B$ and $\varepsilon$). 
This means that the central region and the primordial fragmentation regions have been already formed 
to this time instant. 
The matter in all these regions is in the quark-gluon phase.  
In contrast to high-energy scenarios (at the top RHIC and LHC energies) 
a large fraction of the baryon charge is stopped in the center region. 
The central baryon-rich fireball subsequently expands. This expansion predominantly is of the 1D nature 
at $\sqrt{s_{NN}}>$ 10 GeV. It pushes out the baryon charge to the peripheral regions. 
The primordial 
fragmentation fireballs and the expanding central fireball temporarily keep their identity, see 
energy density at $t=$ 4 fm/c in Fig. \ref{fig2}. 
Later on, at $t\gsim$ 8 fm/c, the primordial 
fragmentation fireballs join with the pushed-out matter of the central fireball. 
These fireballs really join rather than merge, as seen from the last (right) column of 
Fig. \ref{fig2}. The white area inside the freeze-out contours is free of the matter of expanding 
central fireball and is solely occupied by the primordial fragmentation fireballs.

The fine structure of the evolving system along the beam axis ($\eta_s$, $x=y=0$)
is displayed in Fig. \ref{fig3} for central ($b=$ 2 fm) Au+Au collisions at several
collision energies, including 62 GeV. As seen, the numerics at 62 GeV indeed is not 
quite good---large numerical fluctuations take place during all evolution period. 
The proper baryon density 
and  the longitudinal rapidity 
   \begin{eqnarray}
   \label{y_z}
   y_z = \frac{1}{2} \ln\left(\frac{1+v_z}{1-v_z}\right),
   \end{eqnarray}
where $v_z$ is the $z$-component of the collective velocity (\ref{u-tot}) along the beam axis, are displayed 
in Fig. \ref{fig3} for the first-order-transition EoS. 
The crossover results are very similar to those presented in Fig. \ref{fig3}. 
The time instants 
when the equilibration first occurs are marked by the ``equil.'' label. 
As seen, the equilibration occurs later at lower collision energies. 
At earlier time instants, i.e. at $t=2$ fm/c for 11.5 GeV and $t=1$ fm/c for 19.6 GeV, 
the absence of the equilibration is seen already from the velocity profile. 
A good estimate for the equilibration time 
at these energies within the 3FD model 
is $t_{\rm{equil.}} \sim 2 \Delta t_{\rm{pass}}$, i.e. 
the doubled time during which two Lorentz-contracted nuclei (of $R$ radius) pass each other
moving in the opposite directions with the speed of light, 
$\Delta t_{\rm{pass}} \sim  4 m_N R/\sqrt{s_{NN}}$. 
At time instants 
when the equilibration first occurs, the side bumps in the baryon density move in opposite 
directions with velocities close to the speed of light and thus indeed are the primordial
fragmentation fireballs. 
The boundary between these primordial fragmentation fireballs and the central one is also seen 
from the longitudinal velocity profile in Fig. \ref{fig3} even when the fragmentation  
density bumps are not well resolved as at 11.5 GeV.

The central equilibrated fireball is initially produced in the state of the expansion. 
At higher collision energies $\sqrt{s_{NN}}>$ 10 GeV, the central fireball 
undergoes predominantly 1D expansion along the beam direction. 
The matter, and in particular the baryon charge, is pushed out to the periphery of this central fireball, 
i.e. closer to the primordial fragmentation regions, as it usually happens in the 1D expansion. 
At later time instants  the pushed-out matter of the central fireball
continue to move to higher $|\eta_s|$, i.e. accelerate, while
the primordial fragmentation fireballs stay approximately 
at the same space-time rapidity only slightly shifting to higher $|\eta_s|$ because of the
pressure exerted by the pushed-out matter on them. 
This is most clearly seen at the energy of 19.6 GeV in Fig. \ref{fig3}. 
The primordial fragmentation fireballs join with central contributions  
because of the counter expansion of these fragmentation and central fireballs,
see Fig. \ref{fig2}. 
Therefore, the final fragmentation regions consist of primordial fragmentation fireballs  
and baryon-rich regions of the central fireball pushed out to peripheral rapidities.

To gain an impression of the 
baryon charge accumulated in the primordial fragmentation and central regions
we calculate the  
baryon number as $\int dx\;dy\;dz\; n_B/\sqrt{1-v^2}$. 
Under the assumption of 1D expansion, 
which is a good approximation at 
$\sqrt{s_{NN}}>$ 10 GeV, 
the $dxdy$ integration can be 
considered independent of $z$. 
The borders between these regions are determined by wiggles in  
the longitudinal velocity profile, as mentioned above. 
Thus integrating the $n_B$ distribution at the time instants (marked by ``equil.'')
when the equilibration first occurs
we arrive at the following estimate of the fraction of the baryon charge in the central fireball and 
primordial fragmentation regions, see Table \ref{tab:1}. 
\begin{table}[htb]
\begin{ruledtabular}
  \begin{tabular}{|c|cccc|}
$\sqrt{s_{NN}}$ [GeV]              & 11.5 & 19.6 & 39 & 62   \\
central fireball                   & 96\%& 85--90\%& 65-75\%& $\sim$30-50\%  \\
fragmentation fireballs            & 4\%& 10--15\%& 25--35\%& $\sim$50-70\%  \\
  \end{tabular}
\caption{Fraction of the baryon charge in the central fireball and primordial fragmentation fireballs 
right after the initial equilibration in central ($b=$ 2 fm) Au+Au collisions at various 
collision energies. 
}
\label{tab:1}
\end{ruledtabular}
\end{table}
Note that this estimate at 62 GeV is very approximate because of unstable numerics. 
The above estimate has been done for both the first-order-transition and crossover EoS's. 
When the results for these Eos's do not coincide, they are hyphenated 
in Tab. \ref{tab:1}. 
The crossover EoS
predicts slightly larger baryon-charge fraction in the central fireball.

The fraction of the initially equilibrated matter accumulated in the central fireball is 100\% 
at 7.7 GeV. This matter does not undergo strong 1D expansion along the beam 
direction. Therefore, the high baryon density in the midrapidity survives until the freeze-out, 
as it is seen from  Fig. \ref{fig1a}. With the collision energy rise the 
fraction of the initially equilibrated central fireball gradually drops. 
However, it amounts to $\sim$40\% even at the collision energy of 62 GeV. 
With the collision energy rise 
the observable region, i.e. that at the feeze-out stage,  of the high baryon density
gradually moves to fragmentation regions, i.e. to peripheral rapidities, 
while the midrapidity region becomes increasingly baryon-charge depleted.

At higher collision energies $\sqrt{s_{NN}}>$ 10 GeV,
the central part of the system gets frozen out
at the later stage, see panels at $t \geq$ 10 fm/c in Fig. \ref{fig2},
 while the fragmentation regions continue 
to evolve being already separated in the configuration space. 
This longer evolution of the fragmentation regions is because of relativistic time dilation 
caused by the high-speed motion of the fragmentation regions with respect to the central region. 
Therefore, their evolution time is relativistically elongated in the c.m. frame of colliding nuclei
and, e.g., at 39 GeV, lasts $\approx 40$ fm/c. 
At lower collision energies, 
$\sqrt{s_{NN}}<$ 10 GeV, 
the single fireball survives until the very end of the freeze-out.

As seen from Figs. \ref{fig2} and \ref{fig3}, at BES RHIC energies the friction forces 
mainly govern formation of the initially equilibrated state for the further hydrodynamic evolution. 
These friction forces result in a certain interplay between the incomplete baryon stopping and 
the subsequent almost 1D hydrodynamic expansion of the stopped matter. 
The dominance of the incomplete stopping is expected only at top RHIC energies. 
In more detail the nonequilibrium stage of the collisions was analyzed in Ref. \cite{Ivanov:2016hes}
including the entropy production and the 3FD dissipation. 
At the expansion stage of the collision the friction forces provide a moderate 
dissipation that can be interpreted in terms of the shear viscosity \cite{Ivanov:2016hes}.  
The friction in the QGP was fitted to reproduce experimental data at midrapidity 
at BES RHIC energies and in wider rapidity range at SPS energies. 
If the baryon diffusion is incorporated into the hydrodynamics, e.g., like in Ref. \cite{Denicol:2018wdp}, 
it would certainly modify the final midrapidity baryon density. 
If the incorporation of the baryon diffusion gives a better reproduction of the experimental data, 
it may entail changes in the QGP friction and, in its turn, in properties of the 
initially equilibrated state.

\section{Summary}
\label{Summary}

In the present paper 
we estimated the baryon and energy densities 
reached in the fragmentation regions in central Au+Au collisions 
at BES RHIC energies within the 3FD  model.  

It is shown that 
a considerable part of the baryon charge is stopped in the central fireball. 
Even at 39 GeV, approximately 70\% of the total baryon charge turns out to be 
stopped. 
The fraction of the baryon charge stopped in the central fireball decreases with 
collision energy rise, from 100\% at  7.7 GeV to 70\% at 39 GeV. 
A tentative calculation at the energy of 62 GeV results in $\sim$40\% of the stopped 
baryon charge. 
At higher collision energies $\sqrt{s_{NN}}>$ 10 GeV 
the final fragmentation regions are formed from not only primordial fragmentation fireballs,  
i.e. the baryon-rich matter traversed the interaction region, but also of the 
matter of the central fireball pushed out to peripheral rapidities
because of 1D expansion of this central fireball.

The highest initial baryon densities of the equilibrated matter, $n_B/n_0 \approx$ 10, 
are reached in the central region of colliding Au
 nuclei at $\sqrt{s_{NN}}=$ 20--40 GeV. 
These highest densities evolve to quite moderate freeze-out baryon densities 
at the midrapidity because the central baryon 
matter is pushed out to peripheral regions by almost 1D hy-
drodynamic expansion.
Therefore, consequences of these high initial baryon densities 
can be observed only in the fragmentation regions of colliding nuclei in 
experiments at the LHC in the fixed-target mode \cite{Brodsky:2012vg}. 
The highest midrapidity baryon density at the freeze-out is achieved at $\sqrt{s_{NN}}\approx$ 8 GeV,
which approximately agrees with the result of the 
analysis of midrapidity hadron yields within the statistical model  \cite{Randrup:2006nr,Randrup:2009ch}.

All the above features of the collision dynamics are consequences of the strong friction in the QGP, 
i.e. when the counter-streaming regime takes place in the deconfined phase. 
As has been mentioned above, this friction is completely phenomenological, it was  fitted to reproduce 
observables at BES RHIC energies. This fit suggests that the transition into the QGP 
at the stage of interpenetration of colliding nuclei makes the system more opaque.  
It is consistent with jet quenching---if the system is opaque for the jets, it also should be opaque for the counter-streaming baryon flows. What is the mechanism of this counter-streaming opaqueness is still a question. 
It can be the same mechanism as that of the jet quenching. 
If applied to the counter-streaming regime, this mechanism can be associated with the Weibel instability
\cite{Ivanov:1987ef,Mrowczynski:2016etf} that enhances the counter-streaming stopping because of the 
radiation of soft gluons similarly to the radiation in the hadronic phase due to the Weibel instability
\cite{Ivanov:1988dh}.  
Alternatively, it can be due to formation of strong color fields between the leading
partons \cite{Mishustin:2001ib,Mishustin:2006wd}. 
These fields may also enhance baryon stopping as compared to its estimate 
based on hadronic cross-sections \cite{Sat90}.

\vspace*{3mm} {\bf Acknowledgments} \vspace*{2mm}

This work was carried out using computing resources of the federal collective usage center ``Complex for simulation and data processing for mega-science facilities'' at NRC ``Kurchatov Institute'' 
(subvention of the Ministry of Education and Science of the Russian Federation under agreement RFMEFI62117X0016),
http://ckp.nrcki.ru/.
Y.B.I. was supported by the Russian Science
Foundation, Grant No. 17-12-01427.
A.A.S. was partially supported by  the Ministry of Education and Science of the Russian Federation within  
the Academic Excellence Project of 
the NRNU MEPhI under contract 
No. 02.A03.21.0005. 

\end{document}